\documentclass[12pt]{article}

\usepackage{a4wide,epsfig,feynmp}

\newcommand{\gsim}{\raisebox{-3pt}{$\,\stackrel{\textstyle >}{\sim}\,$}}

\newcommand{\lcvec}[3]{\left[\;#1\;,\;#2\;,\;#3\;\right]}

\def\nn{\nonumber}

\def\mev{\,{\rm MeV}}
\def\gev{\,{\rm GeV}}

\def\vk{{\bf k}_\perp{}}

\newcommand{\q}{{\rm q}}
\renewcommand{\d}{{\rm d}}
\renewcommand{\u}{{\rm u}}

\renewcommand{\b}{{\rm b}}

\newcommand{\qbar}{\overline{\rm q}}
\newcommand{\dbar}{\overline{\rm d}}
\newcommand{\ubar}{\overline{\rm u}}

\newcommand{\bbar}{\overline{\rm b}}

\newcommand{\psla}{p\kern-1.0ex/}

\newcommand{\da}{distribution amplitude}
\newcommand{\wf}{wave function}
\def\beq{\begin{eqnarray}}
\def\eeq{\end{eqnarray}}

\begin{document}
\parindent0pt

\begin{fmffile}{btopipic}

\begin{flushright}
WU B 98-10 \\
hep-ph/9905343 
\end{flushright}

\begin{center}
\vskip 3.5\baselineskip
\textbf{\Large Skewed parton distributions 
for $B\to \pi$ transitions}
\vskip 2.5\baselineskip
Th.~Feldmann 
and P.~Kroll
\vskip \baselineskip
Fachbereich Physik, Universit\"at Wuppertal, D-42097 Wuppertal,
   Germany 
\vskip 2.5\baselineskip

\textbf{Abstract} \\[0.5\baselineskip]
\parbox{0.9\textwidth}{
We investigate the $\b$-$\u$ 
skewed parton distributions (SPDs) for $B\to \pi$ transitions
and determine the contributions from several sources
(overlaps of soft light-cone wave functions,
quark-antiquark annihilations and meson resonances).
The $B\to \pi$ transition form factors, which are
relevant in exclusive semi-leptonic and non-leptonic $B$-decays,
are obtained by integrating the $\b$-$\u$ SPDs over the momentum fraction $x$.
A phenomenological determination of the relevant
parameters allows us to predict the form factors 
and to obtain the branching ratios for semi-leptonic 
$B\to\pi$ decays.}
\vskip 1.5\baselineskip
\end{center}


\section{Introduction}

A good theoretical understanding of 
heavy-to-light meson form factors, which encode the confinement of the
quarks in the hadronic bound states, are of utmost interest. Accurate
predictions of the form factors would permit the determination of the
less well-known Cabbibo-Kobayashi-Maskawa (CKM) matrix elements
from experimental rates of exclusive heavy meson decays. For
instance, in the case of the semi-leptonic
 $B\to\pi$ transitions, on which we focus our
interest in this article, the relevant entry in the CKM matrix is
$|V_{\u\b}|$. Its present value is 0.0035 with an uncertainty of about
0.001 \cite{ale:96,PDG}. 
The form factors for transitions from the $B$
meson to light mesons also form an important ingredient
of the calculation of exclusive non-leptonic $B$ decays, 
e.g.\ for $B\to\pi\pi$.
Thus, not surprisingly, the heavy-to-light form factors attracted the
attention of theoreticians, and many articles have been devoted to
their investigation. The theoretical approaches utilised in these
articles reach from the quark model \cite{Isgur:1989gb}, 
overlaps of light-cone \wf s
\cite{Wirbel:1985ji,Szczepaniak:1998xj}, perturbative QCD
\cite{szc:90,Dahm:1995ne}, the heavy quark symmetries 
\cite{Kramer:1991vs,ChHQET} and QCD sum rules
\cite{Khodjamirian:1998ji,bal:98}, to name a few. 

In several of these approaches there are two distinct and prominent
dynamical mechanisms: The $B\pi$ resonances which control the form
factors at small recoil and the overlap of meson wave functions
which dominates at large recoil. 
Other mechanisms, like the perturbative one, provide only small
corrections. The crucial problem arises then, how to match these two
contributions at intermediate recoil. In this article we are proposing
a new approach 
which is based on the concept of
generalised or, as frequently termed, skewed parton distributions
which has recently been invented in the context of deeply
virtual Compton scattering \cite{mue98}. The SPDs are defined as
non-forward matrix elements of non-local currents. They are hybrid objects 
in this respect
which share the properties of ordinary parton distributions
and form factors. We are going to introduce $\b$-$\u$ SPDs as a
parametrisation of the soft $B\to\pi$ matrix element. 
The chief advantage of
the SPDs for $B\to\pi$ transitions is that they clearly separate
resonance and overlap contribution and thus allow the superposition of
both the contributions in an unambiguous way. This SPD approach may
constitute an important step forward towards a unified description of the
$B\to\pi$ form factors at small and large recoil, although we are aware
that there is still a number of open questions to be answered 
before a satisfactory and complete 
description of the $B\to\pi$ transition form
factors has been achieved.  

Our paper is organised as follows:
In Sect.~2 we present the basic definitions and the kinematics.
In Sect.~3 we introduce the $\b$-$\u$ SPDs and discuss 
contributions to them from various sources. From the SPDs 
we calculate the $B\to\pi$ form factors as functions of the momentum transfer,
$q^2$. The results are presented in Sect.~4 together with a comparison
to other results, an assessment of their theoretical uncertainties and
an evaluation of the semi-leptonic $B\to\pi$ decay rates. In
this section we also check our form factors against
the unitarity bounds derived in Ref.\ \cite{Mannel:1998kp}.
The summary is presented in Sect.~5.

\section{Kinematics}

To be specific we consider the semi-leptonic decay
$\bar{B}^0\to \pi^+\ell^-\bar{\nu}_l$; all our results can
straightforwardly be adapted to other $B \to \pi$ transitions.
The form factors for $\bar{B}^0 \to \pi^+$ transitions 
are frequently defined by (see e.g.\
\cite{Wirbel:1985ji,Khodjamirian:1998ji,Mannel:1998kp})  
\beq
\langle \pi^+; p' | \bar{\u}(0) \gamma_\mu {\b}(0) | \bar{B}^0; p\rangle
        &=&
           F_+(q^2) \, \left(p_\mu+p'_\mu - \frac{M_B^2-M_\pi^2}{q^2} \, q_\mu
                                               \right)  \nn \\[0.3em] 
        &+& F_0(q^2) \, \frac{M_B^2 - M_\pi^2}{q^2} \, q_\mu \,,
\label{current-alt}
\eeq
where $q=p-p'$ and $M_B$ ($M_\pi$) is the $B$ ($\pi$) mass. The form
factors defined in (\ref{current-alt}) are subject to the kinematical
constraint $F_+(0)=F_0(0)$.
For our purpose of investigating the SPDs for $B\to \pi$ transitions it is
more convenient to use the alternative
covariant decomposition
\begin{equation}
\langle \pi^+; p' | \bar{\u}(0) \gamma_\mu {\b}(0) | \bar{B}^0; p\rangle =
F^{(1)}(q^2) \, p'_\mu  + F^{(2)}(q^2)\,\left(q_\mu - \frac{q^2} {M_B^2} p_\mu\right)\,.
\label{current}
\end{equation}
The two sets of form factors are related by
\beq
\label{eq:formrel}
F_+ &=& \frac12\,\left( F^{(1)} - \frac{q^2} {M_B^2} F^{(2)}\right) \,,\\
F_0 &=& \frac12\,\left(1-\frac{q^2} {M_B^2-M^2_\pi}\right)\, F^{(1)}
          + \frac{q^2} {2M_B^2}\frac{M_B^2 + M_\pi^2}{M_B^2 - M_\pi^2}
            \,\left(1-\frac{q^2} {M_B^2+M^2_\pi}\right)\, F^{(2)} \,.\nn
\eeq
At $q^2=0$ the form factors $F_+$ and $F_0$  are
solely determined by the form factor $F^{(1)}$.
Most convenient for the calculation of the $B\to\pi$ transition form
factors in terms of SPDs is a frame of reference where the hadron momenta
are collinear to each other; this frame may be viewed as a
generalisation of a Breit frame. We introduce 
light-cone coordinates $v^\pm = (v^0 \pm v^3) /\sqrt{2}$ and 
${\bf v}_\perp = (v^1, v^2)$ for any four-vector $v$ and use
component notation $v= [v^+,v^-,{\bf v}_\perp]$. Defining the
so-called skewedness parameter by 
\begin{equation}
\zeta = \frac{q^+}{p^+} = 1 - \frac{p'^+}{p^+}\,, 
\end{equation}
we can write the $B$ and $\pi$ momenta in our frame of reference as
\begin{equation}
p = \lcvec{p^+}{\frac{M_B^2}{2 p^+}}{{\bf 0_\perp}}\,, \quad\quad
p'= \lcvec{(1-\zeta) p^+}{\frac{M_\pi^2}{2 p^+ \, (1-\zeta)}}{{\bf 0_\perp}}\,.
\label{frame}
\end{equation}
Positivity of the energy of the final state meson implies $\zeta < 1$.
The momentum transfer is given by
\begin{equation}
      q^2 = \zeta M_B^2 \left(1 -  \frac{M_\pi^2}{M_B^2 (1-\zeta)}\right)\,.
\label{q-z}
\end{equation}
The skewedness parameter $\zeta$ covers the interval
$[0,1-M_\pi/M_B]$ 
in parallel with the variation of the momentum transfer from
zero (we neglect the lepton mass here) to $q^2_{\rm max}=
(M_B-M_\pi)^2$ in the physical region of the
$B\to\pi$ transitions. In contrast to the case of form factors in the
space-like region \cite{Diehl:1998kh},
there is no frame for $B\to\pi$ 
transitions in which the skewedness parameter can be chosen to be zero.
In the following we will neglect the pion mass 
in the  calculation of the SPDs and form factors.

For convenience we quote the light-cone components of the current
matrix element (\ref{current}) in the frame of reference (\ref{frame}):
\beq
\langle \pi^+; p' | \bar{\u}(0) \gamma^+ {\b}(0) |\bar{B}^0; p\rangle &=&
                            F^{(1)}(q^2)\, (1-\frac{q^2}{M_B^2})\, p^+\,, \nn\\
\langle \pi^+; p' | \bar{\u}(0) \gamma^- {\b}(0) | \bar{B}^0; p \rangle &=&
            F^{(2)}(q^2) \, (1-\frac{q^2}{M_B^2})\, \frac{M_B^2}{2p^+} \,.
\label{cur-matrix}
\eeq
The matrix elements of the transverse currents are zero.

\section{$\b$-$\u$  skewed parton distributions}

We define the $\b$-$\u$ SPD $\tilde{{\cal F}}^{(1)}_{\zeta}$ by the
non-forward matrix elements
\begin{equation} 
\int \frac{{\d}z^-}{2\pi}\, e^{ixp^+z^-}\, \langle \pi^+; p' |
     \bar{\u}(0) \gamma^+ {\b}(z^-)| \bar{B}^0; p \rangle
                     = (1-\zeta)\, \tilde{{\cal F}}^{(1)}_{\zeta}(x,q^2)\,,
\label{spd-def}
\end{equation}
where $x=k^+/p^+$ is
the fraction of plus-components of the $\b$-quark and $B$-meson
momenta. 
The second SPD, $\tilde{{\cal F}}^{(2)}_{\zeta}$, is analogously
defined  with $\gamma^+$ being replaced by $\gamma^-$,
see also Eq.~(\ref{cur-matrix}). 
In the frame of reference chosen by us the momentum transfer and
the skewedness parameter are related to each other by Eq.\ (\ref{q-z}),
$q^2=\zeta M_B^2$. This relation makes the $q^2$ variable in
$\widetilde{{\cal F}}_{\zeta}^{(i)}$ redundant. For the ease of
notation  we will, therefore, omit it in the following. 

\begin{figure}[t]
\begin{center}
\unitlength0.8cm
\parbox[b]{7cm}{a)
\fmfframe(1,1.0)(1,1.0){
\begin{fmfgraph*}(6,4)
\fmfpen{thick}
\fmfleft{Q1,P1}\fmfright{Q2,P2}\fmftop{P3}
\fmf{dashes,tension=1.5}{Q1,v1} \fmf{dashes,tension=1.5}{v2,Q2}
\fmf{phantom}{v1,v3} \fmf{phantom}{v3,v2} 
\fmffreeze
\fmfv{label=$B(p)$}{Q1}\fmfv{label=$\pi(p')$}{Q2}
\fmfv{label=$q$}{p3}
\fmf{phantom}{v1,v2}
\fmffreeze
\fmfi{phantom,label=:,la.si=right,la.di=0.035w}
  {vpath (__v1,__v3) shifted (thick*(0,3.5)) }
\fmfi{phantom,label=:,la.si=right,la.di=0.035w}
  {vpath (__v3,__v2) shifted (thick*(0,3.5)) }
\fmfi{plain,label=$k{}\phantom{'}$,label.side=left}
  {vpath (__v1,__v3) shifted (thick*(0,3.5))}
\fmfi{plain,label=$k'$,label.side=left}
  {vpath (__v3,__v2) shifted (thick*(0,3.5))}
\fmfi{plain}
  {vpath (__v1,__v2) shifted (thick*(0,-3.5))}
\fmfshift{(thick*(0,3.5))}{v3}
\fmf{photon,tension=1.8}{p3,v3} \fmf{phantom,tension=1.8}{p3,P3}
\fmfdot{v3}{}\fmfblob{.12w}{v2}\fmfblob{.12w}{v1}
\end{fmfgraph*}}
\\
b)
\fmfframe(1,1.0)(1.0,1.0){
\begin{fmfgraph*}(6,4)
\fmfpen{thick}
\fmfleft{Q1,P1}\fmfright{Q2,P2}\fmftop{P3}
\fmf{dashes,tension=1.5}{Q1,v1} 
\fmf{dashes,tension=1.5}{v2,Q2}
\fmf{phantom}{v1,v3}
\fmf{phantom}{v3,v2} 
\fmffreeze
\fmf{phantom}{v1,v2}
\fmffreeze
\fmf{plain,left=0.5,label=$k{}\phantom{'}$}{v1,v4}
\fmf{plain,right=.4}{v1,v4}
\fmfv{label=$k'$,la.a=-40,la.di=0.06w}{v4}
\fmf{photon,tension=2.0}{p3,v4} \fmf{phantom,tension=2.0}{p3,P3}
\fmfv{label=$B(p)$}{Q1}\fmfv{label=$\pi(p')$}{Q2}
\fmfv{label=$q$}{p3}
\fmffreeze
\fmfi{phantom,label=:,la.si=right,la.di=0.05w}
  {vpath (__v1,__v2) shifted (thick*(0,3.5)) }
\fmfi{plain}
  {vpath (__v1,__v2) shifted (thick*(0,2)) }
\fmfi{plain}
  {vpath (__v1,__v2) shifted (thick*(0,-3.5))}
\fmfdot{v4}\fmfblob{.12w}{v1}\fmfblob{.12w}{v2}
\end{fmfgraph*}}}
\hskip2em
c)
\parbox[b]{5cm}{\fmfframe(1,1.0)(1.0,1.0){
\begin{fmfgraph*}(4,4)
\fmfpen{thick}
\fmfleft{Q1,P1}\fmfright{Q2,P2}\fmftop{p3}
\fmf{dashes,tension=1.8}{Q1,v0} 
\fmf{dashes,tension=1.8}{v0,Q2}
\fmffreeze
\fmf{dashes,tension=1.,label=$B^*$}{v0,v1}
\fmfv{decor.shape=square,decor.f=empty,decor.size=.15w}{v0}
\fmf{phantom}{v1,v2}
\fmf{photon,tension=1.5}{v2,p3}
\fmfv{label=$B(p)$}{Q1}\fmfv{label=$\pi(p')$}{Q2}
\fmfv{label=$q$}{p3}
\fmffreeze
\fmf{plain,label=$k{}\phantom{'}$,label.side=left,left=.5}{v1,v2}
\fmf{plain,label=$k'$,label.side=right,right=.5}{v1,v2}
\fmfdot{v2}\fmfblob{.18w}{v1}
\end{fmfgraph*}}}
\end{center}
\caption{\label{fig1} Overlap (a), annihilation (b) and 
  resonance (c) contributions to $B\to \pi$ transitions. The dots
  indicate that any number of spectators may contribute. }
\end{figure}
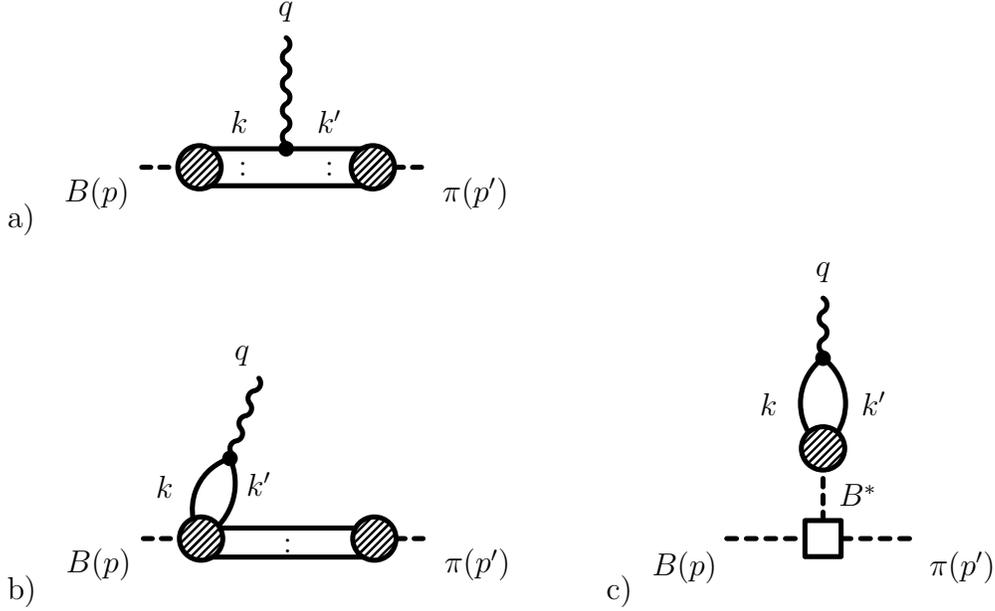

Depending on the value of $x$, the SPDs describe different physical
situations \cite{mue98}: For $1\geq x\geq \zeta$
a $\b$ quark with momentum fraction $x$ is taken out of the $B$
meson and a ${\u}$ quark
carrying a momentum fraction $x'=k^{+}{}'/p^{+}{}'$ (with respect to
the final state meson) is inserted back, turning the $B$ meson into a pion
(see Fig.~\ref{fig1}a)\footnote{The momenta of the active $\b$ and
$\u$ quarks read in our frame of reference
\begin{equation}
k=\lcvec{xp^+}{\frac{m_{\b}^2+k^2_\perp}{2xp^+}}{\vk}\,, \quad\quad
k'=\lcvec{(x-\zeta)p^+}{\frac{m_{\u}^2+k^2_\perp}{2(x-\zeta)p^+}}{\vk}\, .
\end{equation}
}.
This part of the SPDs will be modelled as 
overlaps of $B$ and $\pi$ light-cone \wf s.
For $0\leq x<\zeta$ the $B$ meson emits a ${\b}\ubar$ pair and the remaining
partons form the pion (see Fig.~\ref{fig1}b). According to Brodsky
and Hwang \cite{Brodsky:1998hn} this contribution can be described by
non-diagonal light-cone \wf{} overlaps for $n+2\to n$ parton
processes \footnote{
The existence of this contribution has been stressed  by Sawicki
\cite{saw:92} some time ago.}.
In addition, as pointed out by Radyushkin \cite{Radyushkin:1998bz},
$B\pi$ resonances contribute to the SPDs in that region
(see Fig.~\ref{fig1}c). 
These considerations lead to the following decomposition of the 
$\b$-$\u$ SPDs in the interval $0\leq x \leq 1$ 
\begin{equation}
\widetilde{{\cal F}}_\zeta^{(i)}(x)
     = \theta(x-\zeta)\widetilde{{\cal F}}_{\zeta\,{\rm ove}}^{(i)}(x)\,+\,
       \theta(\zeta-x)\left[ \widetilde{{\cal F}}_{\zeta\,{\rm ann}}^{(i)}(x)
        \, +\, \widetilde{{\cal F}}_{\zeta\,{\rm res}}^{(i)}(x) \right]
                                                                      \,,    
\label{step}
\end{equation}
where the three parts of the SPDs labelled {\sf ove}, {\sf ann}
 and {\sf res} refer to
the contributions from Fig.~\ref{fig1} a), b) and c), respectively. 
The relative importance of the overlap contribution to the SPDs 
on the one side and the sum of annihilation and 
resonance one on the other side, change with the momentum transfer as
a consequence of the relation (\ref{q-z}). 
At large recoil, $q^2 \simeq 0$, the
annihilation and resonance parts do not contribute while 
they dominate at small recoil, $q^2\simeq q^2_{\rm max}$. 
We stress that the superposition (\ref{step}) is controlled by the
momentum transfer or the skewedness parameter $\zeta$
in an unambiguous way, i.e.\ there is no danger of 
double counting. The $\b$-$\u$ SPDs exist in a third region of the
variable $x$,
namely for $-1+\zeta \leq x <0$ where they describe the situation that a
$\b$ quark with a negative momentum fraction is emitted from the
$B$ meson and $\u$ quark is absorbed. Re-interpreting a quark
with a negative momentum fraction as an antiquark with a positive
fraction, one finds that the region $-1+\zeta \leq x <0$ describes the
emission of a $\bbar$-quark and the absorption of a $\ubar$ one.
This re-interpretation implies the relation
\begin{equation}
\widetilde{{\cal F}}_\zeta^{(i)\,\bbar -\ubar}(x) = 
           - \widetilde{{\cal F}}_\zeta^{(i)\,\b -\u}(\zeta - x)
\label{sym}
\end{equation}
for the SPDs ($x\geq \zeta$). By way of exception we here quote the
quark-flavour labels. Since the probability of finding a
$\b\bbar$ sea-quark pair in the $B$ meson is practically zero, 
$\widetilde{{\cal F}}_\zeta^{(i)}(x) \simeq 0$ in the
region  $-1+\zeta \leq x <0$ to a very high degree of accuracy.

By comparison of (\ref{cur-matrix}) and (\ref{spd-def}) one finds the
reduction formula
\begin{equation}
   F^{(i)}(q^2)\,=\, \int^1_0 {\d}{x} \, \widetilde{{\cal
                                              F}}^{(i)}_{\zeta}(x)
\label{fff}
\end{equation}
for $i=1,2$. The range of the $x$ integration is
restricted to the interval $[0,1]$ since contributions from $\bbar$
quarks or, in other words, from negative momentum fractions are absent
in the form factors.  

As already mentioned, we describe the overlap part of the SPDs by
light-cone \wf s for the $B$ and the $\pi$ mesons. 
To begin with we consider the valence Fock states of the $B$ and $\pi$
mesons. The corresponding light-cone \wf s, $\Psi_B$ and $\Psi_\pi$,
respectively, provide the overlap contribution  
\begin{equation}
\widetilde{{\cal F}}_{\zeta\,{\rm ove}}^{(1)}(x)\,=\,
             \frac{2}{1-\zeta} \, \int \frac{{\d}^2 \vk}{16\pi^3}
   \;\Psi_\pi^{*}(x'=\frac{x-\zeta}{1-\zeta}, \vk)\;
               \Psi_B (x,\vk) \,, 
\label{par-spd}
\end{equation}
where $\vk$ is the intrinsic
transverse momentum of the $\b$ ($\u$) quark with respect to the $B$
($\pi$)-meson momentum. As a consequence of the collinearity of the
two meson momenta in our frame of reference the transverse momentum
in the argument of the $\pi$ \wf{} is the same as in the $B$ \wf{},
while the longitudinal momentum fraction is shifted. 

For the pion valence Fock state \wf{} we take a simple Gaussian ansatz
\begin{equation}
\Psi_\pi (x,\vk) = \frac{\sqrt6}{f_\pi} \, 
        \exp\left[- \frac{1}{8\pi^2 f_\pi^2} \, \frac{\vk^2}{x \, (1-x)}\right]
\label{pi-wf}
\end{equation}
with the associated asymptotic \da{}  
\begin{equation}
\phi_\pi^{\rm AS}(x) = 6 x (1-x) \,. 
\label{pi-da}
\end{equation}
$f_\pi$ (=132\mev) is the usual pion decay constant. The pion's
transverse size parameter is fixed
by the chiral anomaly  to $(2\sqrt{2} \pi f_\pi)^{-1}$ \cite{BHL}. 
The \wf{} (\ref{pi-wf}) being normalised to 0.25 at a scale of 1 \gev{}, 
has been tested against experiment and found to work satisfactorily in
many hard exclusive reactions involving pions (cf.\ \cite{raulfs} for
instance). It is also supported by recent QCD sum rule results (cf.\
\cite{braun} for instance), by a study of power corrections
\cite{akh:98} and by the instanton model \cite{goeke}.

For the $\b \qbar$ \wf{} of the $B$ meson we use a slightly modified
version of the 
Bauer-Stech-Wirbel (BSW) function \cite{Wirbel:1985ji} which has been shown
to be useful in weak decays
\footnote{A Gaussian as in (\ref{pi-wf}) for the $B$ meson has
theoretical deficiencies in the formal limit $M_B\to\infty$ and is,
therefore, in conflict with the heavy quark effective theory \cite{koe:93}.}
\begin{equation}
\Psi_B(x,\vk) =  
\frac{ f_B}{2\sqrt6} \, \phi_B(x) 16 \pi^2 a_B^2\; \exp{[-a_B^2
                                                           \vk^2]}\,,
\label{B-wf}
\end{equation}
where the \da{} is given by
\begin{equation}
\phi_B(x) =  
{\mathcal N} \, x \, (1-x) \, 
 \exp\left[-a_B^2 \, M_B^2 \, \left(x - x_0)\right)^2 \right]\,.
\label{phiBansatz}
\end{equation}
The \da{} $\phi_B$ exhibits a pronounced peak, its
position is approximately at $x\simeq x_0=m_{\b}/M_B$. For a $\b$-quark
mass, $m_{\b} $, of 4.8 \gev{} the value of $x_0$ is 0.91. 
This property of the $B$-meson
\da{} parallels the theoretically expected and
experimentally confirmed behaviour of heavy meson fragmentation
functions. The constant ${\cal N}$ in Eq.\ (\ref{phiBansatz}) is fixed
by the condition  
\begin{equation}
\int_0^1 \phi_B(x)\, dx = 1 \,. 
\label{da-norm}
\end{equation}
For the $B$-meson decay constant $f_B$ we take a value of 180~\mev{}
which is supported by recent lattice gauge theory analyses \cite{alex}. 
The only remaining free parameter in the $B$-meson \wf s
(\ref{B-wf}) is the transverse size parameter, $a_B$, which we fix by 
normalising the $B$-meson's valence Fock state probability to unity.
This leads to a value of 1.51 \gev$^{-1}$ for $a_B$ if a value of 4.8 
\gev{} is chosen for the $\b$-quark (pole) mass \cite{kuh:98}. The
parameter $\bar{\Lambda}$, given by the $B$-meson and $\b$-quark 
mass difference, acquires a value of 480\mev. 
The constant ${\cal N}$ in Eq.~(\ref{phiBansatz}) then takes a
value of 54.7. The maximum of the \da{} $\phi_B(x)$ is located at 
$x_{\rm max}=0.86$. We checked that our final results only mildly 
depend on variation of the parameters $m_b$ and $f_B$ and of the 
probability of the $B$ meson's valence Fock state.

Performing the trivial $\vk$ integration in (\ref{par-spd}), we find
\begin{equation}
\widetilde{{\cal F}}_{\zeta\,{\rm ove}}^{1}(x) = 
           \, 8\pi^2 f_B f_\pi\, a_B^2\; 
         \frac{(x-\zeta)\, (1-x)}
             {8\pi^2 f_\pi^2 a_B^2\, (x-\zeta)\, (1-x)\, +\, (1-\zeta)^2} 
             \; \frac{\phi_B(x)}{1-\zeta} \,.
\label{eq:overlap}
\end{equation} 
We see that $\widetilde{{\cal F}}_{\zeta\,{\rm
ove}}^{1}(x) \propto (x-\zeta)$ for $x\to\zeta$ and $\zeta$ fixed. 
In the formal limit $M_B\to\infty$, this SPD behaves as
$M_B^{-3/2}$. 
  
Of course, the result (\ref{eq:overlap}) can easily be translated to
other choices of the pion and $B$-meson wave functions. The numerical
results will not change significantly as long as \da s
are used  which
are close to the asymptotic one in case of the pion and 
strongly peaked at large $x$ in case of the $B$ meson.

{}From the \wf s (\ref{pi-wf}) and (\ref{B-wf})
one may also calculate the overlap part of the SPD
$\widetilde{{\cal F}}_{\zeta\,{\rm ove}}^{2}$ in full analogy to
(\ref{par-spd}). However, due to additional $k_\perp$ factors,
arising from the matrix elements of the $\gamma^-$ current
between the light-cone helicity spinors, 
$\widetilde{{\cal F}}_{\zeta\,{\rm ove}}^{2}$ is 
power-suppressed to order $\bar{\Lambda}/M_B$ at least as compared to
$\widetilde{{\cal F}}_{\zeta\,{\rm ove}}^{1}$ and, hence, neglected.

In principle, the overlap parts of the SPDs receive
contributions  from all Fock states. 
The generalisation of the overlap representation (\ref{par-spd}) to
higher Fock states is a straightforward application of the methods
outlined in \cite{Diehl:1998kh}. Using suitably generalised wave
functions for the higher Fock states (cf.\ \cite{Diehl:1998kh}), one
can show that the higher Fock state contributions to $\widetilde{{\cal
F}}_{\zeta\,{\rm ove}}^{1}$ are very small and can be
neglected. It is not only the tiny probabilities of the higher
$B$-meson Fock states which is responsible for this fact. Even more important
for the suppression of these contributions is a conspiracy of the
factor $(1-x)^{n(N)}$ appearing in the $N$-particle Fock state
contribution to the SPD $\widetilde{{\cal F}}_{\zeta\,{\rm ove}}^{1}$
and the strongly peaked shape of the $B$-meson wave function.
Here, $n(N)$ is a positive integer increasing with $N$
\cite{Diehl:1998kh}. Since $x_0 =1-\bar{\Lambda}/M_B$ one may regard the
contribution of the 
$N$-particle Fock state as a power correction $(\bar{\Lambda}/M_B)^{n(N)}$
to (\ref{eq:overlap}). Thus, to a high degree of accuracy, the
restriction to the valence contribution suffices for the overlap part
of the $\b$-$\u$ SPDs.

In order to estimate the annihilation part of the SPDs we can restrict
ourselves again to the parton process with the minimal number of partons
participating, namely the process $\b\dbar \u\ubar \to \dbar u$, 
and we are going to show that this contribution is negligibly
small, too. Numbering the $\b$ quark by 1 and $\ubar$ by 4
and noting that the momentum $q$ is shared by the $\b$ and the $\ubar$
quark, one finds the conditions $x_4=\zeta -x_1$ and $\vk{}_4
=-\vk{}_1$ in our frame of reference defined by Eq.\ (\ref{frame}). In
combination with momentum conservation this leads to the relations $x_3=1-\zeta
-x_2$ and $\vk{}_3 =-\vk{}_2$ for the momentum fractions and
transverse momenta of the additional $\dbar$ and $\u$ quarks. With
these results in mind one arrives 
at the following overlap contribution \cite{Brodsky:1998hn}
\begin{equation}
 \widetilde{{\cal F}}_{\zeta\,{\rm ann}}^{1} (x) =
            \frac{2}{1-\zeta}\, \int_0^{1-\zeta}
                   \, {\d}x_2 \int \frac{{\d}^2\vk{}_1 {\d}^2\vk{}_2}{(16\pi^2)^3}
               \,  \Psi^*_\pi (x'_2=\frac{x_2}{1-\zeta}, \vk{}_2)
                  \, \Psi_{B,4} (x_i, \vk{}_i) \,,
\end{equation}
where $\Psi_\pi$ is the pion valence Fock state wave function
(\ref{pi-wf}) and $\Psi_{B,4}$ the four particle wave function of the
$B$ meson. Generalising the \wf{} (\ref{B-wf}), (\ref{phiBansatz}) in a
straightforward fashion to the four-particle case, we find 
\begin{equation}
\widetilde{{\cal F}}_{\zeta\, {\rm ann}}^{(1)}(x) \propto
                 (1-\zeta)^2 (\zeta-x) 
                          \exp{\left[-a_B^2 M_B^2 (x -x_0)^2\right]} \,,
\end{equation}
i.e.\ the annihilation contribution to $\widetilde{{\cal F}}_{\zeta}^{(1)}$ is
exponentially damped except for
$\zeta \gsim x_0$ (see Eq.\ (\ref{step}))
and $x \simeq x_0$. This
region, however, is suppressed by the factors $(1-\zeta)^2$ and
$\zeta-x$. Thus, the annihilation contribution is very small and can
safely be neglected. Similar arguments hold for $\widetilde{{\cal
F}}_{\zeta\, {\rm ann}}^{(2)}$. 

For the resonance contribution (see Fig.~\ref{fig1}c) we concentrate
on the resonance that is closest to the physical decay region, i.e.\
on the $B^{*-}$ vector meson. {}From the Lorentz structure
of the $BB^*\pi$ vertex \cite{ChHQET} we infer 
\beq
\widetilde{{\cal F}}_{\zeta\, {\rm res}}^{(1)}(x) &=&
                           \frac{f_{B^*} \, g_{B B^*\pi}}{M_{B^*}} \,
                           \left(M_{B^*}^2 - \frac12 \zeta M_B^2 \right) \,
                           \frac{\phi_{B^*}(x/\zeta)}
                           {M_{B^*}^2 - \zeta \, M_B^2} \nn \ , \\[0.3em]
\widetilde{{\cal F}}_{\zeta\, {\rm res}}^{(2)}(x) &=&
                           \qquad - \frac12 \, M_B^2 \,
                           \frac{f_{B^*} \, g_{B B^*\pi}}{M_{B^*}} \,
                           \frac{\phi_{B^*}(x/\zeta)}
                           {M_{B^*}^2 - \zeta \, M_B^2}\,.
\label{eq:pole}
\eeq
The valence Fock state of the $B^{*-}$ resonance consists of a $\b$
and a $\ubar$ quark with an associated wave function similar 
to Eq.\ (\ref{B-wf}). Since the transverse parton momenta, defined
with respect to the $B^*$ momentum $q$, are integrated over, only the 
$B^*$ distribution amplitude, $\phi_{B^*}(y)$, remains for which one
may, for instance, apply the same ansatz as for the $B$ meson. Its 
explicit form is irrelevant for the transition form factors as we will
see below. The 
argument of the $B^*$ distribution amplitude, $x/\zeta$, equals the
momentum fraction $k_+/q_+$ the $\b$-quark carries w.r.t.\ the
$B^*$ meson.
In the numerical analysis to be
discussed below we take
$ f_{B^*} \ g_{BB^*\pi} = 20 \, f_B$ which is compatible with  
a recent QCD sum rule analysis \cite{Khodjamirian:1998ji}.
The coupling constant of the $BB^*\pi$ vertex is related to
a parameter $g$ in an effective Lagrangian in which the chiral and
heavy quark symmetries are built in \cite{ChHQET}, by
$g_{BB^*\pi}= 2 M_B \, g/f_\pi (1 + {\cal O}(\Lambda_{\rm QCD}/M_B))$. 
{}From this it follows that for $q^2 \simeq q^2_{\rm max}$ the SPD
$\widetilde{{\cal F}}_{\zeta\, {\rm res}}^{(1)}$ 
scales as $f_{B} \, g \, M_B/M_\pi \propto M_B^{1/2}$.
This scaling law is in accordance with the one for the corresponding
$B \to \pi$ form factors which has been found 
from the heavy quark limit of QCD in a model-independent 
way \cite{Kramer:1991vs}.

\begin{figure}
\begin{center}
\epsfclipon
\psfig{file=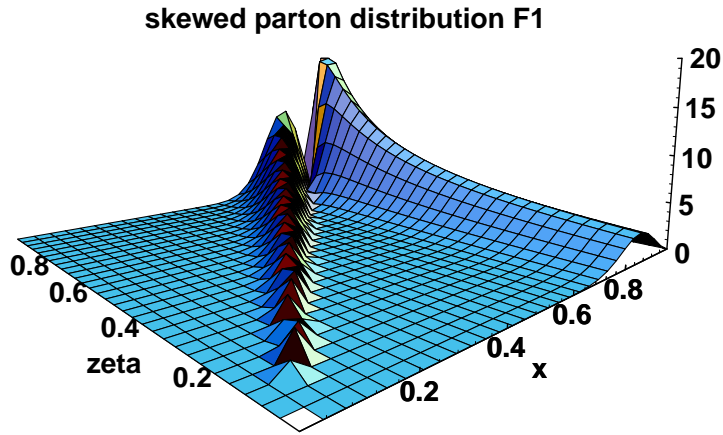, bb= 65 245 545 560, width=10cm}
\end{center}
\caption{
The SPD $\widetilde{{\cal F}}_\zeta^{(1)} (x)$ vs. $x$ and $\zeta$.}
\label{fig2a}
\end{figure}

Putting all this together we obtain the numerical results for the
$\b$-$\u$ SPD   
$\widetilde{{\cal F}}_{\zeta}^{(1)}$ displayed in Fig.~\ref{fig2a}.
Due to the characteristic features of the $B$-meson
\da{}, the overlap contribution (\ref{eq:overlap})
to $\widetilde{{\cal F}}_{\zeta}^{1}$ exhibits a bump at
$x\simeq x_0$ provided $\zeta$ is smaller than $ x_0$. That bump becomes more
pronounced if $\zeta$ approaches $x_0$. 
The  resonance contribution (\ref{eq:pole}) provides the ridge at $x
\protect \simeq x_0 \, \zeta$ where the 
$B^*$ \da{} is large.
The resonance contribution generates a similar ridge in
$\widetilde{{\cal F}}_{\zeta}^{(2)}$ while 
the overlap contribution to it is zero in our model.

Comparing the properties of both, the overlap and the resonance parts
of which our SPDs consist,
we see that they are continuous at the border points $x=0$ and
$x=\zeta$ while their derivatives do not exist there. Hence, our
$\widetilde{{\cal F}}_\zeta^{(i)} (x)$ are non-analytic at the border
points, a property that, according to Radyushkin
\cite{Radyushkin:1998bz}, the SPDs should possess.

\section{$B \to \pi$ form factors}

While the form factor decomposition (\ref{current}) is appropriate for
the investigation of the SPDs, the form factors $F_+$ and $F_0$ are
more suitable in applications to decay processes. We therefore
refrain from discussing the form factors $F^{(i)}$, $i=1,2$ and present
numerical results for $F_{+,0}$ only. 
The overlap contributions to the latter form factors
 are obtained from Eqs.\ (\ref{eq:overlap}), (\ref{step}) and
(\ref{fff}) by numerical integration and insertion of the resulting form
factor $F^{(1)}$ into Eq.\ (\ref{eq:formrel}). The resonance
contribution can be found along the same lines. In this case the $x$
integration is trivial since it only applies to the $B^*$ \da{} and, as
a change of variables reveals,
this integral is just the normalisation (\ref{da-norm}).
Hence, one obtains for the resonance contribution to the form factor $F_+$ 
\beq
F_{+, {\rm res}}(q^2) &=& \frac12\, \frac{q^2}{M_B^2} \;
                          \frac{f_{B^*} \, g_{B B^*\pi} M_{B^*}}
                          {M_{B^*}^2 -q^2} \,.
\label{ffp-res}
\eeq
Note that the standard monopole term (see e.g.\ \cite{ChHQET}) is
modified by the factor $q^2/M_B^2$ which implies a $q^2$-dependent
$B^*$ coupling to the $B\pi$ system. That factor arises from our
ansatz (\ref{eq:pole}) in combination with Eq.\ (\ref{step}) and
(\ref{fff}). Since we consider a large range of momentum transfer (with
respect to the meson radii) the appearance of such $q^2$ dependence is
not unreasonable. It forces the resonance contribution to
vanish at $q^2=0$ in concord with the physical interpretation of the SPDs,
see Eq.\ (\ref{step}). At $q^2\simeq q^2_{\rm max}$, on the other
hand, the resonance term (\ref{ffp-res}) is very close to the standard 
monopole term.

Analogously, one finds for the resonance
contribution to the form factor $F_0$ 
\beq
F_{0, {\rm res}}(q^2) &=& \frac{q^2}{M_B^2} \;
                          \left(1 - \frac{q^2}{M_B^2}\right) \,
                          \frac{f_{B^*} \, g_{B^*B\pi}}{2 \, M_{B^*}} \,.
\label{ff0-res}  
\eeq
A pre-factor, 
arising from the combination of  Eqs.~(\ref{eq:formrel}) and
(\ref{eq:pole}), cancels the $B^*$-pole in $F_0$. (We remind the
reader of the fact that $F_0$ refers to a scalar current.) 

In addition to the overlap and resonance contributions the form factors
also receive contributions from perturbative QCD where 
a hard gluon with a virtuality of the order of
$M_B^2$ is exchanged between the struck and the spectator
quark. In Ref.\ \cite{Dahm:1995ne} the perturbative contributions
have been evaluated at large recoil within the modified perturbative
approach in which 
the transverse degrees of freedom are retained and Sudakov
suppressions taken into account. Since in Ref.~\cite{Dahm:1995ne} the same
soft wave functions with Gaussian suppressions of large intrinsic
transverse quark momenta have been applied as here (see Eqs.\
(\ref{pi-wf},\ref{B-wf})), we can make use of the results presented in
\cite{Dahm:1995ne} and add
them to our form factor predictions. At small recoil, $q^2\geq
18$~GeV$^2$, the predictions for the
perturbative contributions cease to be reliable
because of the small virtualities some of the internal off-shell
quarks and gluons acquire in this region.

\begin{figure}[hbt]
\begin{center}
\unitlength0.8cm
\epsfclipon
{\psfig{file=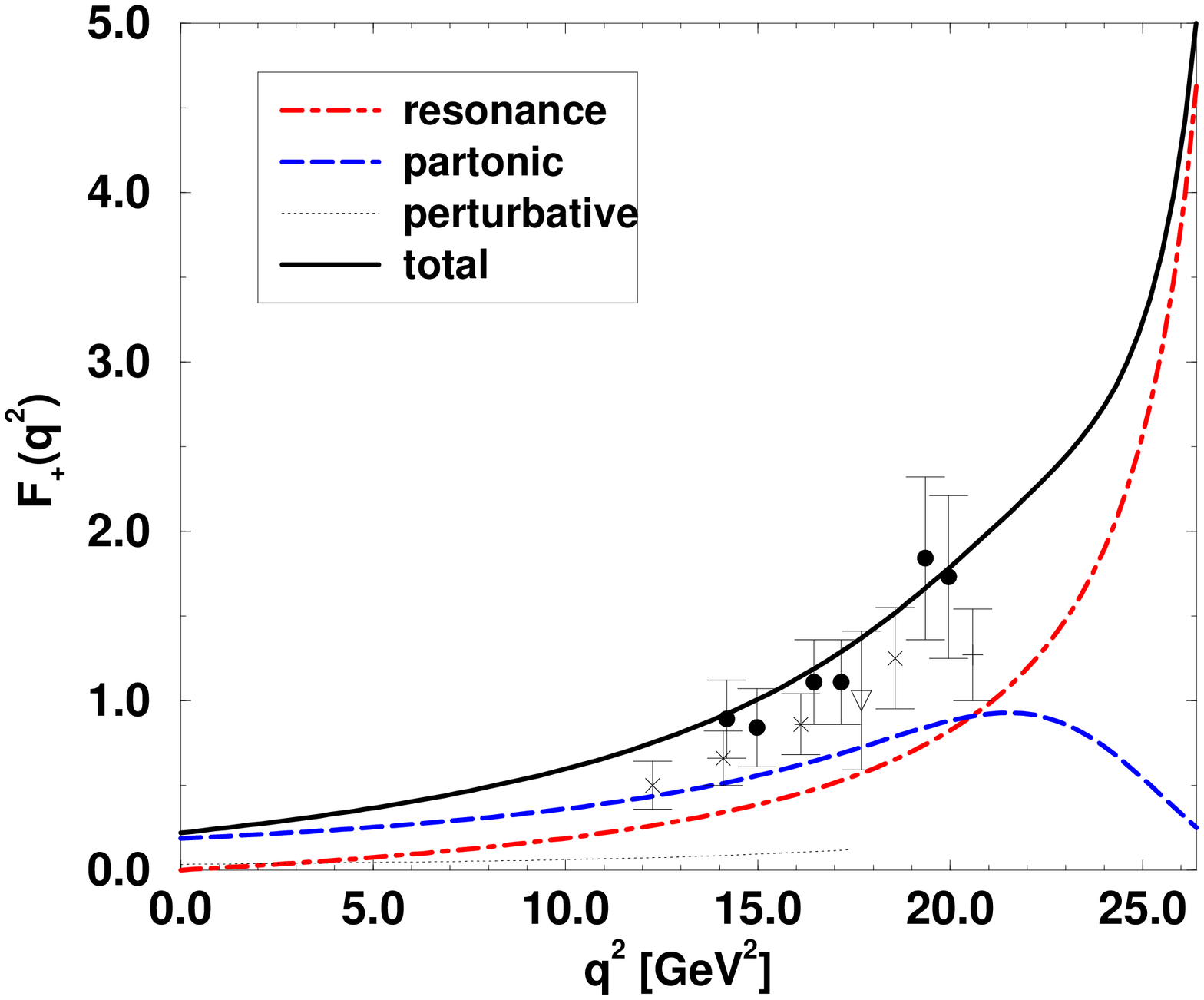, bb=60 70 570 625, 
	width=8.5\unitlength, angle=-90}}
\hskip2em
{\psfig{file=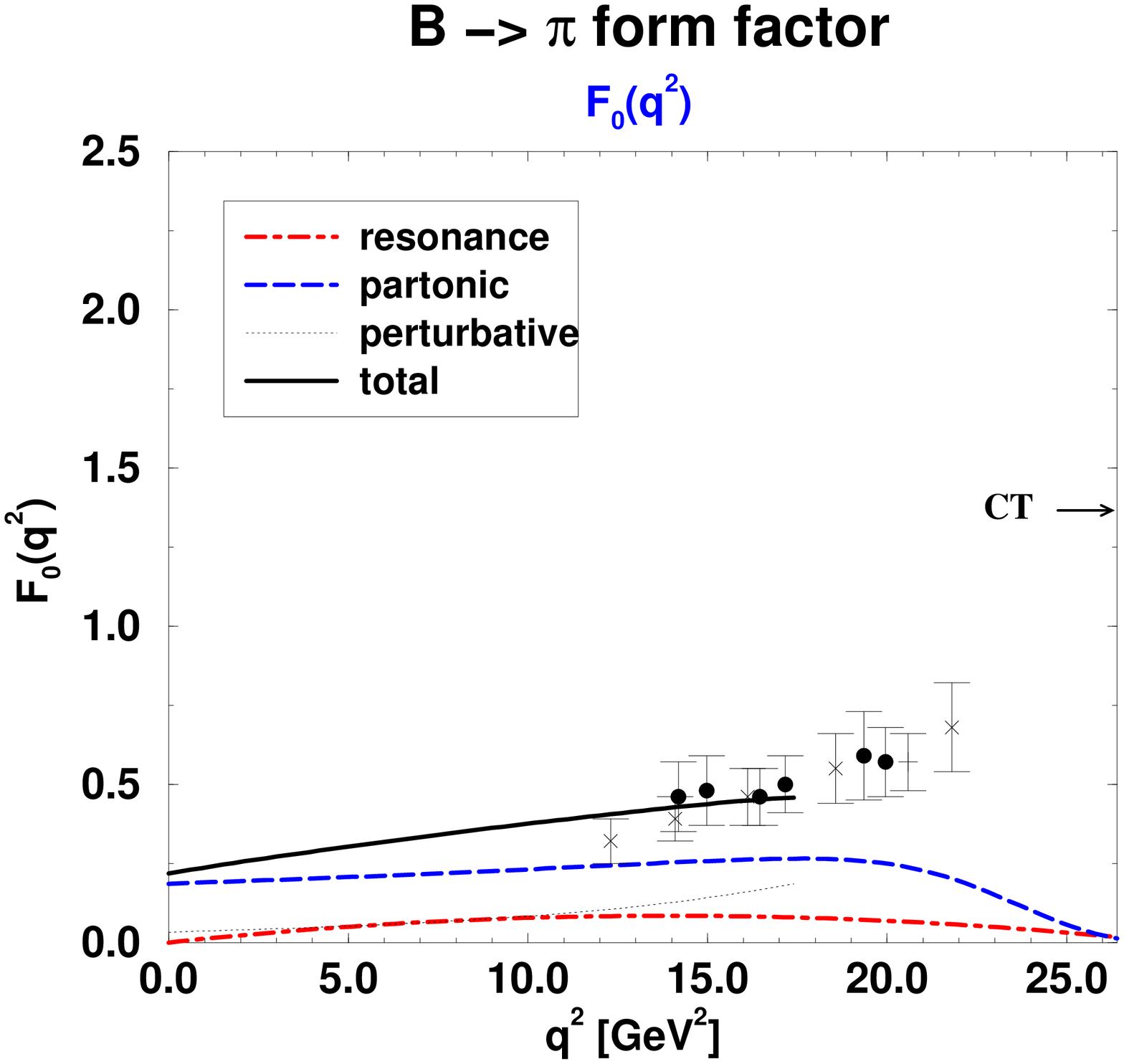, bb=60 70 570 625, 
	width=8.5\unitlength, angle=-90}}
\end{center}
\caption{The form factors $F_+(q^2)$ and $F_0(q^2)$ vs.\ momentum transfer. 
Our predictions (solid lines) for the form factors are decomposed into
resonance, overlap and perturbative contributions. The lattice QCD data, 
taken from Ref.\ \cite{Flynn:1996rc}, are shown for comparison. CT
indicates the Callan-Treiman value, see text.} 
\label{fig3}
\end{figure}

Numerical results for the three contributions, the overlap, the
resonance and the perturbative one, are plotted in Fig.~\ref{fig3}.
In the case of the form factor $F_+$ we observe the dominance of the
overlap contribution at large recoil while the resonance contribution
takes the lead at small recoil. This feature is expected to hold from the
decomposition (\ref{step}).
The perturbative contribution, taken from Ref.\ \cite{Dahm:1995ne}, provides
only a small correction to $F_+$, of the order of $10\%$, at large
recoil and can be neglected at $q^2\simeq q^2_{\rm max}$
as compared to the large resonance contribution. 
Actually, for the numerical analysis the perturbative
contribution to $F_+$ is smoothly continued to zero
for $q^2\geq 18$~GeV$^2$. 
The sum of the three contributions to $F_+$ is in fair 
agreement with the lattice QCD results presented in \cite{Flynn:1996rc}.

Due to the absence of the $B^*$ pole the form factor $F_0$ behaves
differently; it is rather flat over the full range of momentum
transfer. The perturbative contribution makes up a substantial
fraction of the total result for $F_0$ at intermediate momentum
transfer. Since, as we mentioned above, 
it becomes unreliable for $q^2 \gsim 18 \gev^2$ we are not in the
position to predict $F_0$ at large $q^2$. 
A calculation of $F_0$ in
that region would also require a
detailed investigation of the scalar $B\pi$ resonances of which not
much is known at present. Despite of this drawback 
our results for this form factor are also in fair agreement
with the lattice QCD results \cite{Flynn:1996rc} and, in tendency, seem 
to extrapolate to the $B$-sector analogue of the Callan-Treiman value  
\beq
  F_0^{CT}(q^2=q_{\rm max}^2) &=& \frac{f_B}{f_\pi} + {\cal O}(M_\pi^2/M_B^2)
\eeq
which is provided by current algebra in the chiral limit \cite{ChHQET}.

\begin{figure}[hbt]
\begin{center}
\psfig{file=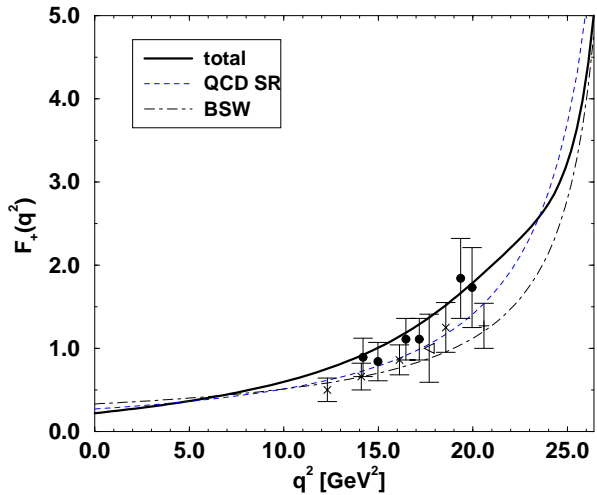, bb=60 70 570 625, width=7cm,angle=-90}
\end{center}
\caption{Comparison of various predictions for the form factor $F_+$.
The solid line represents our result, the dashed one the QCD sum rule
result of Ref.\ \cite{Khodjamirian:1998ji} and the dash-dotted one
the BSW result \cite{Wirbel:1985ji}. The lattice QCD data are taken
from Ref.\ \cite{Flynn:1996rc}.}
\label{fig4}
\end{figure}

In Fig.~\ref{fig4} we compare our results to a few other predictions
of the $B\to \pi$ form factor $F_+$. We first mention the work by
Bauer, Stech and Wirbel \cite{Wirbel:1985ji} in which the form
factor has been calculated from a light-cone \wf{} overlap at $q^2=0$
and the result is used as a normalisation of a pole term. The BSW
model has been applied to exclusive $D$- and $B$-meson decays and
works quite well phenomenologically in many cases. Bauer, Stech and
Wirbel employ a parameterisation of the pion \wf{} which resembles
that of their $B$ \wf{} (see (\ref{B-wf}), (\ref{phiBansatz})) and,
in contrast to us, normalise the pion \wf{} to unity. Doing so they
find a larger overlap and, hence, a larger form factor as we do, see
Fig.~\ref{fig4}. At intermediated momentum transfer, on the other
hand, our predictions for $F_+$ exceed the BSW result as a
consequence of the superposition of resonance and overlap contribution.

Khodjamirian and R\"uckl \cite{Khodjamirian:1998ji} employed QCD
light-cone sum rules for the calculation of the $B\to\pi$ form
factors. In this approach the soft matrix elements are expressed as a
series of collinear terms arising from operators of increasing twist;
actually operators are used up to twist 4. The soft contributions are
supplemented by $\alpha_s$ corrections to the twist-2 contribution
and, at large momentum transfer where the QCD sum rules become
unstable, by the $B^*$ resonance matched to the sum of the other
contributions at $q^2\simeq 16\gev^2$. 
As  Fig.~\ref{fig4} reveals there are similar
deviations between our predictions  and those presented in
\cite{Khodjamirian:1998ji} although to a lesser extend as
in the case of the BSW model. A QCD sum rule analysis of the
$B \to \pi$ form factors has also been attempted by Ball
\cite{bal:98}. Although the results for $F_+$ presented in
\cite{Khodjamirian:1998ji} and \cite{bal:98} agree fairly well with
each other in general, differences in details are to be noticed. 

Of particular interest is the value of the form factor $F_+$ at zero
momentum transfer. It plays an important role in the rates of
the semi-leptonic $B$-meson decays and also in exclusive $B$-decays
into $\pi\pi$ or other pairs of light pseudoscalar mesons. The widths
for the latter processes are calculated on
the basis of a (weak interaction) factorisation hypothesis with
eventual QCD corrections. The quality of that hypothesis is not
well-known. The factorising contribution to the decay
amplitude is proportional to the $B\to\pi$ form factor $F_0$ at
$q^2\simeq 0$. From recent investigations of exclusive
non-leptonic $B$-decays
\cite{ali:1998} we learned that a value of $F_0(0)=F_+(0)$ in the
range of 0.30-0.33 is needed in order to account for the experimental 
decay widths \cite{CLEO} within that approach. Such a large value
cannot easily be accommodated by the models. With the exception of the
BSW model \cite{Wirbel:1985ji} where a value of 0.33 for $F_+(0)$ has
been obtained, most of the other approaches,
e.g.\ \cite{Szczepaniak:1998xj,Khodjamirian:1998ji,bal:98},
provide values within the range of 0.2-0.3 
which are often subject to
substantial uncertainties so that there is no obvious
conflict with the present theoretical understanding 
of non-leptonic $B$ decays. For
instance, in the QCD sum rule approach proposed in Ref.\
\cite{Khodjamirian:1998ji} a value of 0.27 has been found with an 
estimated error of about 0.05. We predict a value of 0.22 for $F_+(0)$
of which an amount of 0.03 originates from the perturbative
contribution. For an assessment of the theoretical uncertainties of
our results we have to consider the following items:\\
i) The overlap contribution is subject to Sudakov suppressions of the
end-point region, $x\to 1$. Since the \wf s we are using,
(\ref{pi-wf}), (\ref{B-wf}) and (\ref{phiBansatz}), already suppress
that region substantially (as compared, for instance, to the ones
used in Ref.~\cite{Szczepaniak:1998xj})
we do not expect the inclusion of the Sudakov
factor to lead to dramatic effects. Moreover, the Sudakov suppressions
are compensated to some extend by ${\cal O}(\alpha_s^2)$ corrections to
the perturbative contributions \cite{Szczepaniak:1998xj}. Thus, we
estimate that the net effect of Sudakov suppression and ${\cal
O}(\alpha_s^2)$ corrections does not exceed $10\%$ of the overlap
contribution to the form factor $F_+$.\\
ii) In the QCD sum rule approach \cite{Khodjamirian:1998ji,bal:98} a
not unimportant contribution to the form factors comes from a
two-particle twist-3 \da{}. That \da{} is constrained by the
vacuum-pion matrix element of the pseudoscalar current being related to
the divergence of the corresponding axial-vector current matrix element
and known to acquire the large value $f_\pi M_\pi^2/(m_\u + m_\d)$
where the $m_\q$ represent current quark masses. 
The implementation of this constraint 
into the light-cone \wf{} approach is somewhat
ambiguous, and we therefore refrain from it in this article.  It
requires the introduction of a pion valence \wf{} component where
quark and antiquark are in opposite helicity states. Such a component
has been discussed in connection with the Melosh transform, see 
e.g.\ \cite{dzi:87}.
By examining several plausible parametrisations of
this \wf{} component we find that its
numerical impact on the overlap is around $10\%$.\\
iii) One may consider deviations of the pion \da{} from the asymptotic
form (\ref{pi-da}). Markedly broader \da s, used for instance in
recent QCD sum rule analyses \cite{Khodjamirian:1998ji,bal:98},
clearly enhance the overlap with the $B$-meson \wf{}. On the other hand,
they are in conflict with the $\pi\gamma$ transition form factor and
the parton distributions of the pion \cite{raulfs}. In order to
examine the bearing of the form of the pion \da{} on the size of the
overlap contribution we allow for a value of $\pm 0.01$ for the
second coefficient, $B_2$, in the Gegenbauer expansion of that
\da{}. Such a value of $B_2$, being still tolerated by the
$\pi\gamma$ transition form factor within the light-cone wave function
approach, leads to a change of  $\pm0.03$ for $F_+(0)$.\\
iv) The uncertainty of the resonance contribution is proportional to
that of the product of coupling constant and the $B^*$ decay constant 
which is about $20\%$.\\ 
Combining these uncertainties with those arising from the input
parameters in our approach ($f_B$, $m_\b$) and the
neglected order $\bar{\Lambda}/M_B$ corrections, we estimate the total
uncertainty of our results for the $B \to \pi$ transition form factors to
be about $20$-$25\%$.

\begin{figure}[hbt]
\unitlength0.8cm
\epsfclipon
\begin{center}
a)
\psfig{file=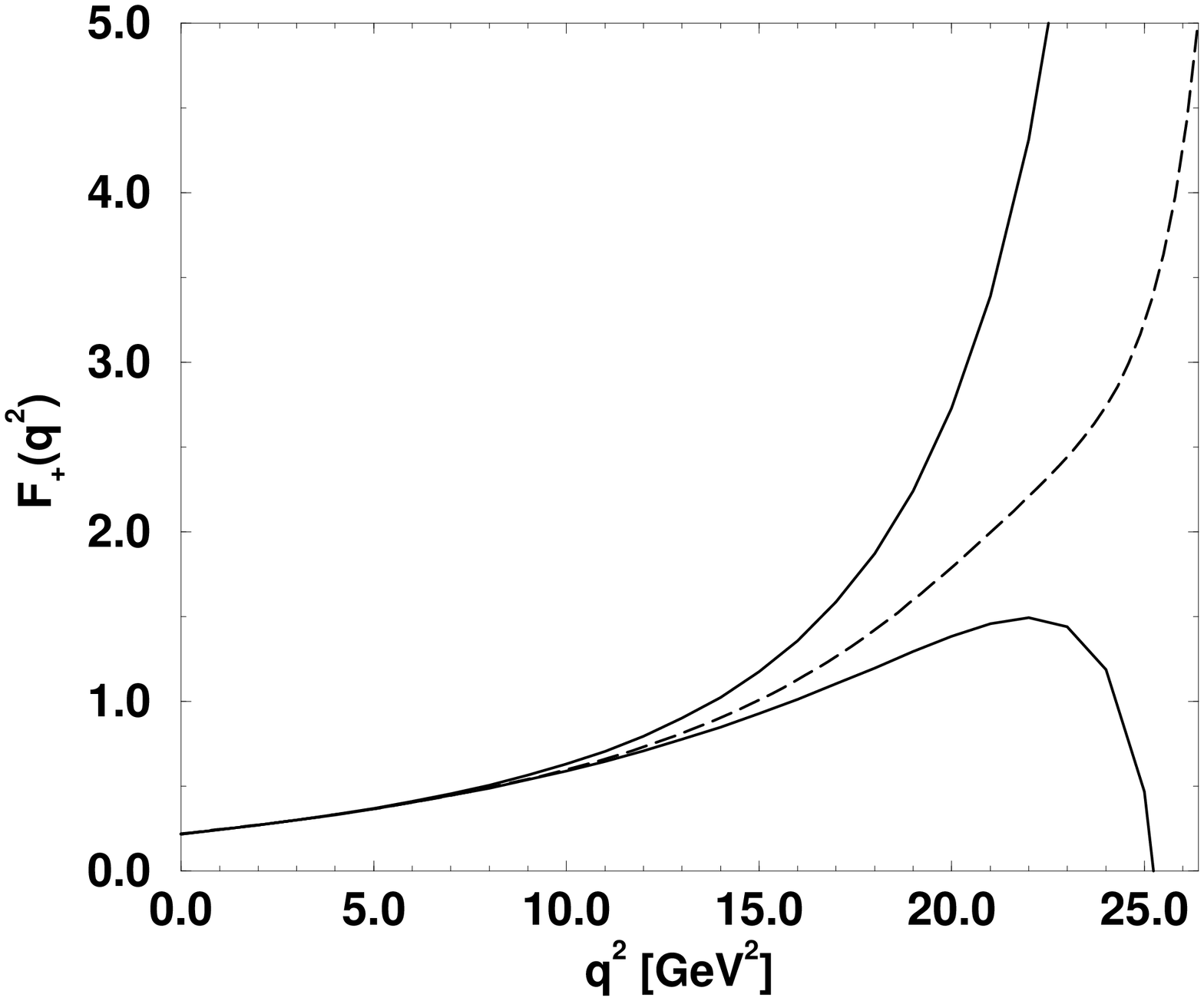, bb=95 108 570 683, 
        angle=-90,width=8\unitlength}
\psfig{file=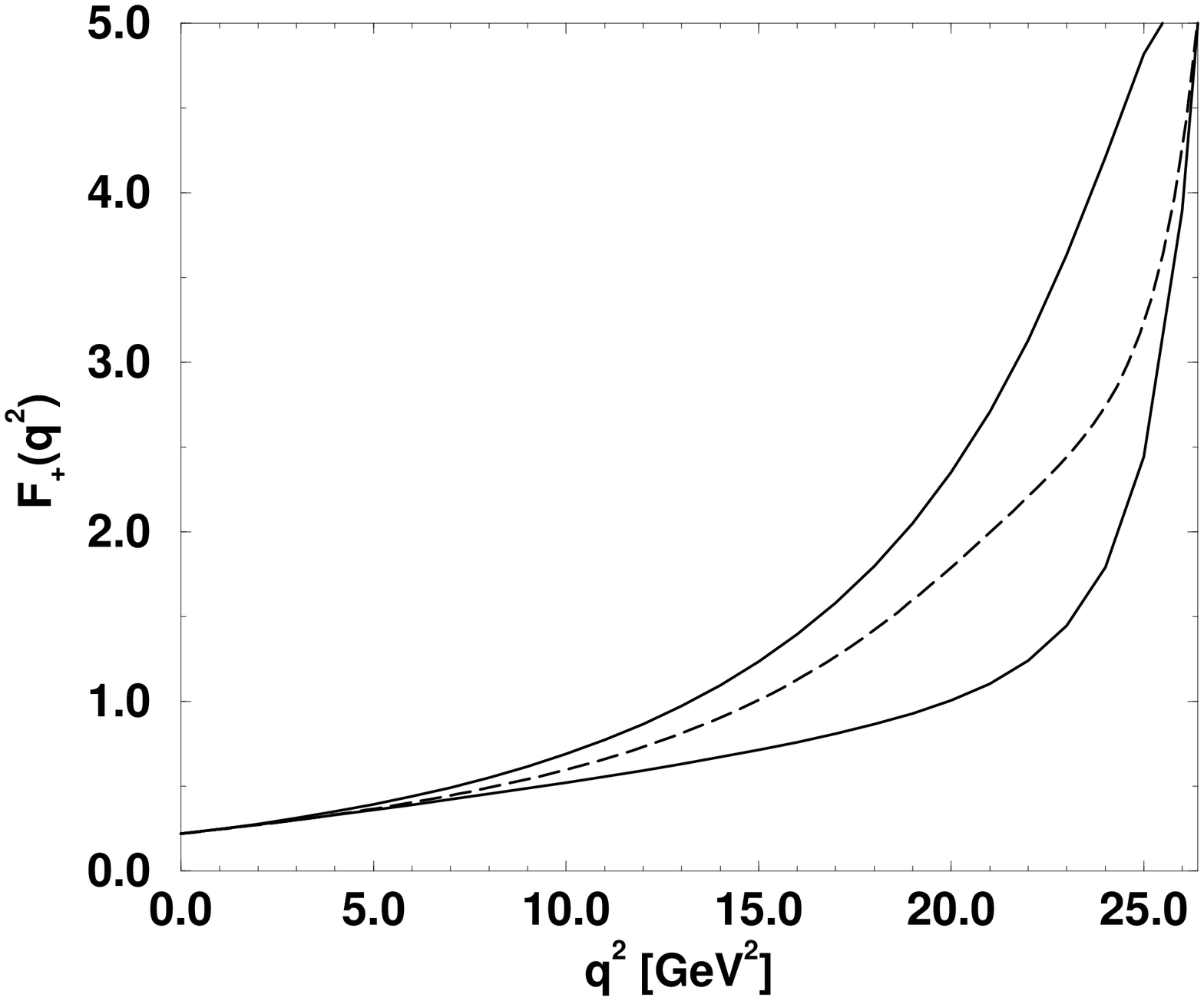, bb=95 35 570 610, 
        angle=-90,width=8\unitlength}
b)
\\
c)
\psfig{file=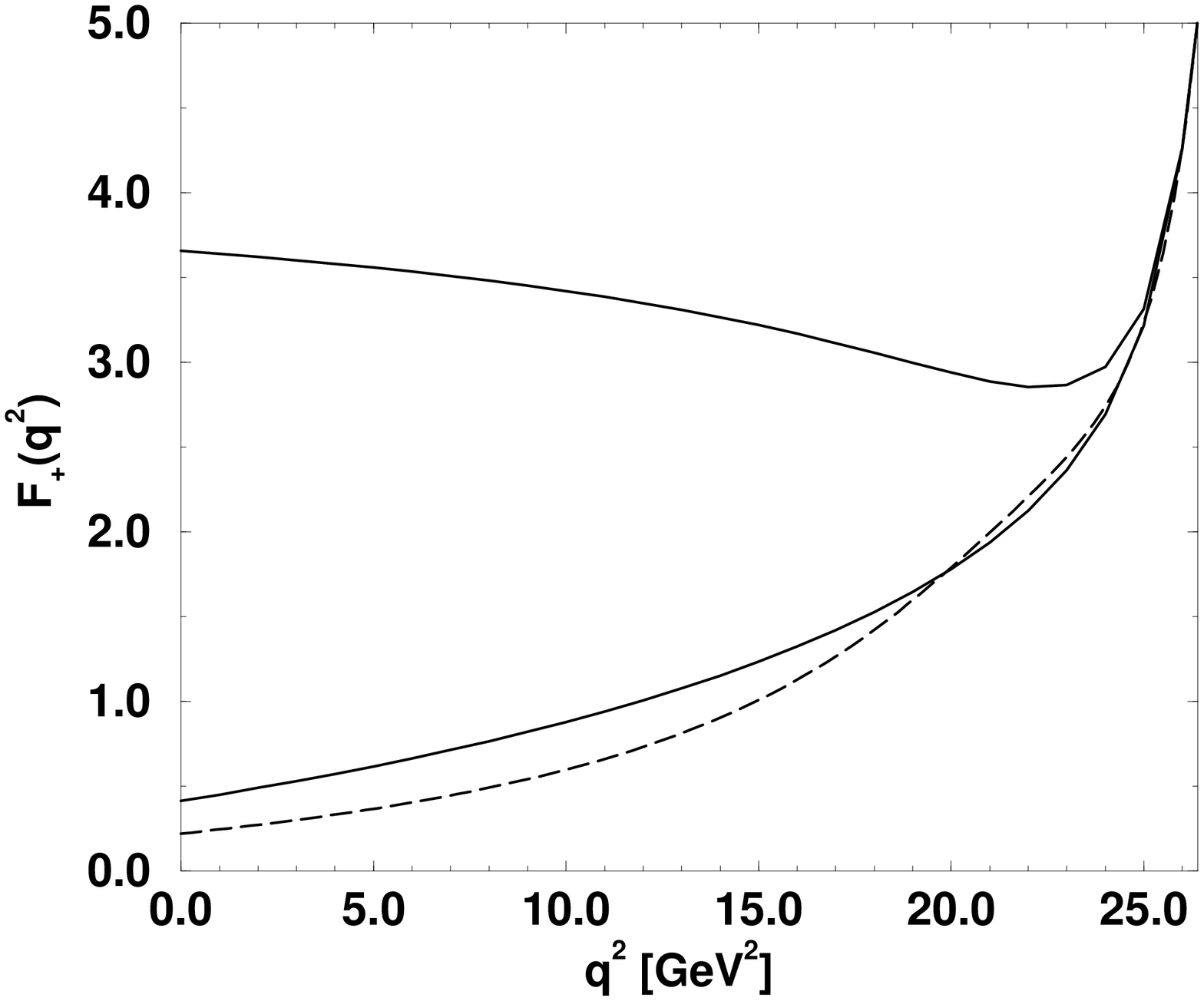, bb=95 108 515 683,
        angle=-90,width=8\unitlength}
\psfig{file=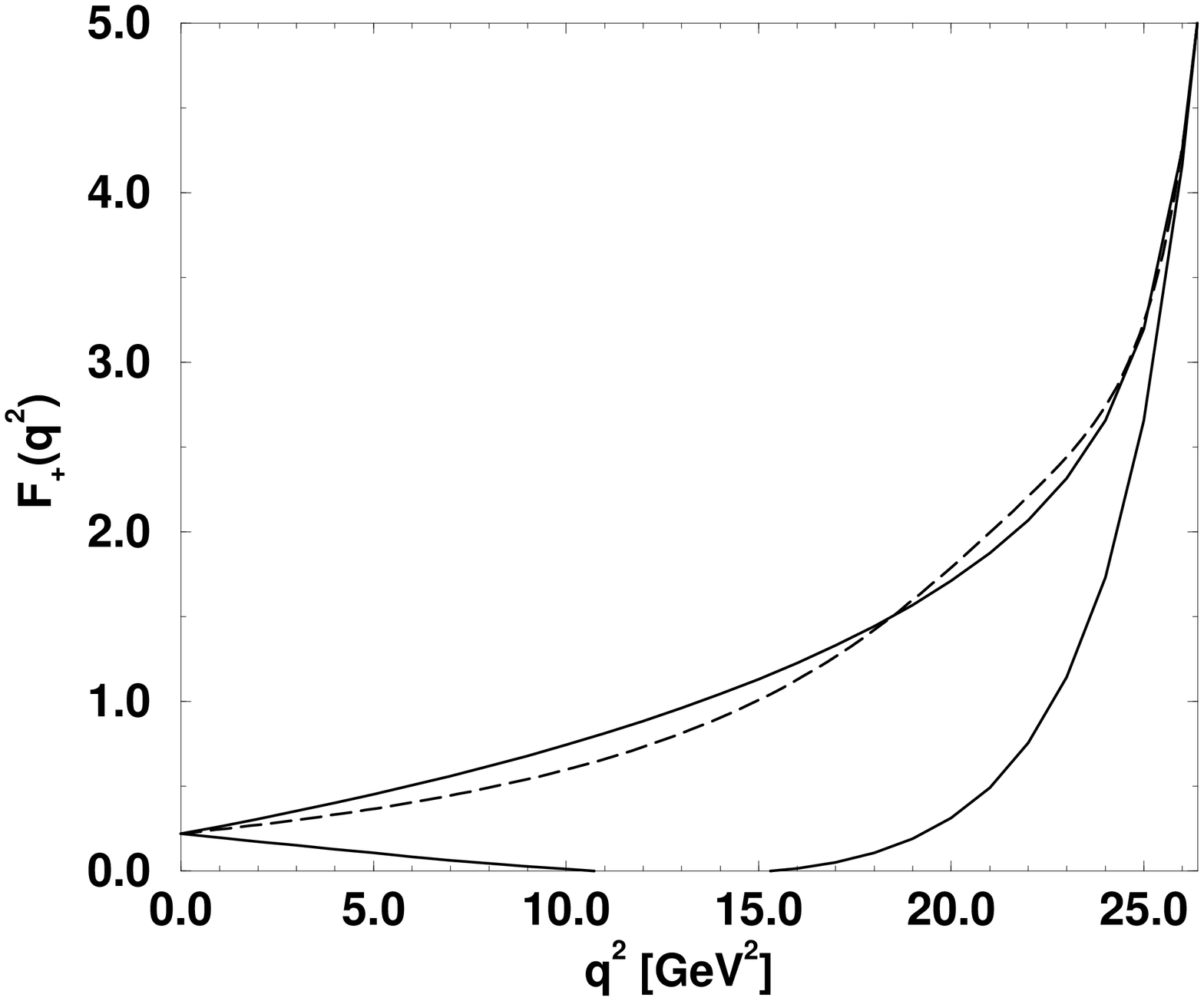, bb=95 35 515 610, 
	angle=-90,width=8\unitlength}
d)
\end{center}
\caption{Testing unitarity. Solid lines: unitarity bounds
\cite{Mannel:1998kp}; dashed lines: our results for $F_+$.
a) The value, slope and curvature of $F_+$ at $q^2=0$ are given;
b) value and slope at $q^2=0$, value at $q^2_{\rm max}$;
c) value, slope and curvature at $q^2_{\rm max}$;
d) value at $q^2=0$, value and slope at $q^2_{\rm max}$.}
\label{fig5}
\end{figure}

Mannel and Postler \cite{Mannel:1998kp} derived model-independent
bounds for the $B\to \pi$ transition form factors from analyticity and
unitarity. Inclusion of the values of the form factors and their
derivatives at minimum and/or
maximum momentum tighten the bounds considerably which
then become a stringent test of the internal consistency of a model
and its compatibility with QCD. 
We submit our form factor $F_+$ 
to this examination and take its values at $q^2=0$
and $q^2=q^2_{\rm max}$ as
well as its first two derivatives  at $q^2=0$ as input. 
The result is plotted in Fig.~\ref{fig5}a) and b).
We observe that our prediction for $F_+$ lies comfortably
within the bounds.

One may also consider bounds for given  slope and curvature
of $F_+$ at $q^2=q^2_{\rm max}$. However, in
contrast to the value
of $F_+(q^2_{\rm max})$ itself which is dominated by the 
$B^*$ pole, the higher derivatives of $F_+$ at small
recoil may be sensitive to corrections from additional
resonances, the treatment of perturbative corrections in
that region etc.
Nevertheless, for the sake of completeness, we plot
the unitarity bounds with
given $F_+'(q^2_{\rm max})$ and $F_+''(q^2_{\rm max})$
in Fig.~\ref{fig5}c) and d).
A mild violation of the bounds is observed.
In view of the systematic and parametric
uncertainties discussed above this is not to be considered
as an inconsistency of our approach.

Let us now turn to the discussion of the semi-leptonic
decay rates $\bar{B}^0\to \pi^+ \ell^- \bar{\nu}_\ell$. 
The differential decay rate is given by 
\begin{eqnarray}
\frac{d \Gamma}{d q^2} &=& \frac{G^2 |V_{\u\b}|^2}{24\pi^3} \,
                         \frac{(q^2-m_l^2)^2\,
                           \sqrt{E_\pi^2-M^2_\pi}}{q^4 M^2_B}\,\nn \\
&& \times  \left\{(1+\frac{m^2_l}{2q^2})\, M_B^2
                      (E_\pi^2-M_\pi^2)\, |F_+(q^2)|^2 + \frac{3m^2_l}{8q^2}\,
           (M^2_B-M^2_\pi)^2\,|F_0(q^2)|^2\right\} \,,
\label{eq:width}
\end{eqnarray}
where $E_\pi=(M^2_B+M^2_\pi-q^2)/(2M_B)$ is the pion energy in the $B$-meson
rest frame. It is important to realize that for light leptons 
the scalar form factor $F_0$ plays a negligible role in the decay rate
since its contribution appears with the square of the lepton mass,
$m_l$. Therefore, the differential decay rates for the light-lepton modes
determine $|V_{\u\b} F_+(q^2)|$. On the other hand, the
heavy-lepton decay mode $\bar{B}^0\to \pi^+{\tau}\bar{\nu}_\tau$ 
offers the possibility of exploring the scalar form factor. 

\begin{figure}[hbt]
\begin{center}
\psfig{file=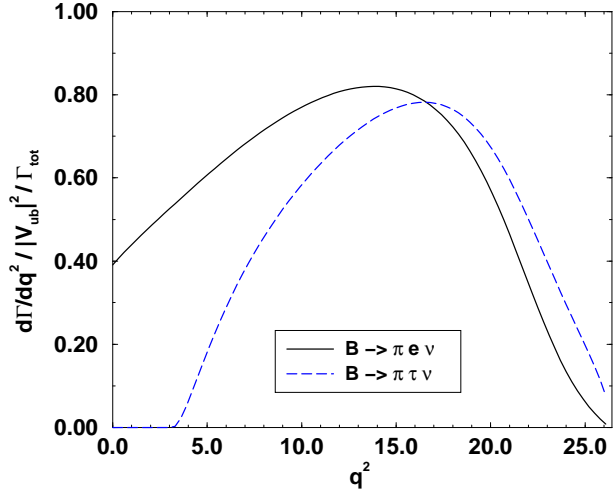, bb=60 70 570 625, width=7cm,angle=-90}
\end{center}
\caption{Predictions for the semi-leptonic 
differential decay widths, divided by
$|V_{\u\b}|^2$ and the total $\bar{B}^0$-meson width, $\Gamma_{\rm
tot}$, vs.\ momentum transfer.
Solid line: $\bar{B}^0\to \pi^+ e\bar{\nu}_e$; dashed line: $\bar{B}^0\to
\pi^+\tau\bar{\nu}_\tau$.} 
\label{fig6}
\end{figure}

Our predictions for the semi-leptonic decay rates into
light or $\tau$ leptons are shown in Fig.~\ref{fig6}. 
For the scalar form factor $F_0$, which becomes important
in the $\tau$ mode,
we use her a simple, smooth interpolation between the CT value at
$q^2=q^2_{\rm max}$ and our results for $F_0$ below $q^2 = 18
\gev^2$. For the branching ratio of the light-lepton modes we find
$$ BR[\bar{B}^0\to \pi^+ e \bar{\nu}_e] \simeq BR[\bar{B}^0 \to \pi^+
\mu \bar{\nu}_\mu] = 1.9 \cdot 10^{-4} 
\cdot \left(\frac{|V_{\u\b}|}{0.0035}\right)^2 .$$ 
The theoretical uncertainty of this prediction, dominated by that of
the overlap contribution, amounts to about
$30\%$. Our result is to be compared with the CLEO measurement
\cite{ale:96}: $(1.8\pm0.4\pm0.3\pm0.2) \cdot 10^{-4}$ 
where the quoted errors refer to the statistical and systematical 
uncertainties and to the model dependence of the CLEO analysis, respectively.

For the $\tau$ channel we obtain
$$ BR[\bar{B}^0\to \pi^+ \tau \bar{\nu}_\tau] = 1.5 \cdot 10^{-4}
\cdot \left(\frac{|V_{\u\b}|}{0.0035}\right)^2.$$
The estimated theoretical error amounts to about $30\%$.
The ratio of both the branching ratios, 
in which the CKM matrix element cancels, amounts to $0.78$ with an
uncertainty of $15\%$.

\section{Conclusions}

We investigated the $\b$-$\u$ SPDs within a light-cone \wf{}
approach. Besides the usual overlap of the $B$ and $\pi$ valence Fock
state \wf s we also considered higher Fock states as well as
annihilation contributions from non-diagonal overlaps and showed that
these contributions provide only small, negligible corrections to the
leading valence term. The $B^*$ resonance is an important and,
at small recoil, dominant contribution and has to be taken into
account for a complete description of the transition form factors. The
chief advantage of the SPD approach is that the skewedness parameter
clearly separates the overlap from the resonance contribution and
both the contributions can be added in an unambiguous way. 
{}From the $\b$-$\u$
SPDs we calculated the $B\to\pi$ transition form factors by means of
reduction formulas. Taking into account the corrections
from perturbative
physics \cite{Dahm:1995ne},
we obtain a reliable predication of
$F_+$ for the entire range of momentum
transfer and for $F_0$ up to about 18\gev$^2$. 
In particular, we obtain a value of
$0.22\pm 0.05$ for the form factors at maximum recoil. This value
appears to be somewhat small if contrasted to the value required in
$B\to \pi\pi$ decays (if the latter process is analysed on the basis
of the factorisation hypothesis) but it is within range of other
theoretical predictions of $F_+(0)$
\cite{Wirbel:1985ji,Szczepaniak:1998xj,Khodjamirian:1998ji,bal:98}.
Generally, our results for the form factors are in fair agreement with
the QCD sum rule result of Khodjamirian and R\"uckl
\cite{Khodjamirian:1998ji} which is, in spirit, very close to the
light-cone \wf{} approach. Our results are in agreement with lattice
QCD data \cite{Flynn:1996rc} and
respect the unitarity bounds derived
in Ref.\ \cite{Mannel:1998kp}, leaving aside mild violations
for cases where the derivatives of $F_+$ at $q^2=q^2_{\rm max}$
are used as input. 

Using our form factors we calculated the differential and total decay
rates for semi-leptonic $B\to\pi$ decays. Our predictions for the
total decay for the process $\bar{B}^0\to\pi^+e\bar{\nu}_e$ is in good
agreement with the recent CLEO measurement \cite{ale:96} if a value of
0.0035 is used for the CKM matrix element $|V_{\u\b}|$.
We stress that the knowledge of $F_+(0)$ is not sufficient for a
prediction of the total decay rates since the $q^2$ dependence of the form
factors is model-dependent.

We finally note that our approach can straightforwardly be applied to
other heavy-to-light meson transition form factors.
At small recoil
the heavy quark symmetries \cite{ChHQET} 
may turn out helpful in fixing parameters. 

\end{fmffile}

\subsection*{Acknowledgement}
We like to thank Boris Postler for the numerical check of the
unitarity bounds and helpful comments.
T.F.\ is supported by {\it Deutsche Forschungsgemeinschaft}\/.



\end{document}